\documentclass[prd,12pt]{revtex4}
\pdfoutput=1 
\usepackage{amsmath}
\usepackage{graphicx}
\usepackage{subfigure}
\usepackage{bm}
\usepackage{slashed}

\def\d{\mathrm{d}}

\begin{document}
\title{Predicting the $B \to \rho$ form factors using AdS/QCD Distribution Amplitudes for the $\rho$ meson}

\author{M. Ahmady}
\affiliation{Department of Physics, Mount Allison University, Sackville, N-B. E46 1E6, Canada}
\email{mahmady@mta.ca,recampbell@mta.ca} 

\author{R. Campbell}
\affiliation{Department of Physics, Mount Allison University, Sackville, N-B. E46 1E6, Canada}
\email{recampbell@mta.ca} 

\author{S. Lord}
\affiliation{D\'epartement de Math\'ematiques et Statistiques, Universit\'e de Moncton,
Moncton, N-B. E1A 3E9, Canada}
\email{esl8420@umoncton.ca}

\author{R. Sandapen}
\affiliation{D\'epartement de Physique et d'Astronomie, Universit\'e de Moncton,
Moncton, N-B. E1A 3E9, Canada
\& \\
Department of Physics, Mount Allison University, Sackville, N-B. E46 1E6, Canada }
\email{ruben.sandapen@umoncton.ca} 

\begin{abstract}
 We use QCD light-cone sum rules (LCSR)  with holographic anti de Sitter/Chromodynamics (AdS/QCD)  Distribution Amplitudes (DAs) for the $\rho$ meson to predict the form factors that govern the leading amplitudes of  the rare radiative $B \to \rho \gamma$ decay
and the semileptonic decay $B \to \rho l \nu$. We test the Isgur-Wise relation between the radiative and semileptonic form factors. We also compute the total width (in units of $|V_{ub}|^2$) for the semileptonic decay $B \to \rho l \nu$ as well as ratios of partial decay widths which are independent of $|V_{ub}|$. 
\end{abstract}


\maketitle

\section{Introduction}
 In a previous paper \cite{Ahmady:2012dy}, we  derived four holographic AdS/QCD DAs for the $\rho$ meson: two twist-$2$ DAs, one for each polarization of the $\rho$, and two twist-$3$ DAs, vector and axial vector, for the transversely polarized $\rho$. We 
used two of them, namely the transverse twist-$2$ DA  in order to compute the next-to-leading order $\alpha_s$ contribution in the leading power  $B \to \rho \gamma$ amplitude and the vector twist-$3$ DA to compute power-suppressed annihilation contributions. The four DAs are all derived from a single AdS/QCD light-front wavefunction for the $\rho$ meson which was shown to generate successful predictions for diffractive $\rho$ electroproduction \cite{Forshaw:2012im}. 

Our goal in this paper is to use both the longitudinal and transverse twist-$2$ AdS/QCD DAs in QCD light-cone sum rules (LCSR) \cite{Ali:1993vd,Ball:1997rj,Ball:1998kk,Ball:2004rg} in order to compute one form factor for the radiative decay $B \to \rho \gamma$  as well as three form factors for the  semileptonic decay 
${B} \to \rho l \nu$. The first decay mode is important for precision tests of the Standard Model and to constrain new physics. The latter decay is useful for extracting the CKM matrix element $|V_{ub}|$.

Having computed the form factors for the semileptonic decay, we are then able to predict the total decay width, $\Gamma$, in units of the CKM matrix element $|V_{ub}|$. We shall also compute observables which are independent of $|V_{ub}|$. Firstly, we predict the ratios of the partial decay widths in different bins of the momentum transfer $q^2$. These partial decay widths have recently been measured by the BaBar collaboration \cite{delAmoSanchez:2010af} for three different bins: $q^2 \in [0,8]~\mbox{GeV}^2$, $q^2 \in [8,16]~\mbox{GeV}^2$ and $q^2 \in [16,20.3]~\mbox{GeV}^2$. Secondly, we shall predict the ratio $\Gamma_L/\Gamma_T$ where $\Gamma_{L}(\Gamma_{T})$ is the partial decay rate to a final state where the $\rho$ is longitudinally (transversely) polarized.  A future measurement of this quantity would serve as a test for our prediction. 

\section{AdS/QCD Distribution Amplitudes}

The AdS/QCD DAs are so-called because they are related to the light-front wavefunction of the $\rho$ meson and the latter wavefunction can be obtained by solving the holographic light-front Schroedinger equation \cite{deTeramond:2008ht,Brodsky:2008kp} for mesons.  In Ref. \cite{Ahmady:2012dy}, we have shown that the twist-$2$ AdS/QCD DAs are given by
\begin{equation}
\phi_{\rho}^\parallel(z,\mu) =\frac{N_c}{\pi f_{\rho} m_{\rho}} \int \d
r \; \mu
J_1(\mu r) [m_{\rho}^2 z(1-z) + m_f^2 -\nabla_r^2] \frac{\phi_L(r,z)}{z(1-z)} 
\label{phiparallel-phiL}
\end{equation}
and
\begin{equation}
\phi_{\rho}^{\perp}(z,\mu) =\frac{N_c m_f}{\pi f_{\rho}^{\perp}} \int \d
r \; \mu
J_1(\mu r) \frac{\phi_T(r,z)}{z(1-z)} 
\label{phiperp-phiT}
\end{equation}
where $\phi_{\lambda=L,T}(r,z)$ is the AdS/QCD light-front wavefunction of the $\rho$ meson given by \cite{Vega:2009zb}
\begin{equation}
\phi_{\lambda} (r,z)= \mathcal{N}_{\lambda} \frac{\kappa}{\sqrt{\pi}}\sqrt{z(1-z)} \exp \left(-\frac{\kappa^2 \zeta^2}{2}\right) \exp\left(-\frac{m_f^2}{2\kappa^2 z (1-z)} \right)
\label{AdS-QCD-wfn}
\end{equation}
with $\zeta=\sqrt{z(1-z)} r$ being the light-front variable\footnote{ $r$ is the magnitude of the transverse separation between the quark and the antiquark  while $z$ is the fraction of the meson light-front momentum carried by the quark.} that maps onto the fifth dimension of AdS space \cite{deTeramond:2008ht}. The AdS/QCD wavefunction given by  Eq. \eqref{AdS-QCD-wfn} is obtained using a quadratic \cite{Brodsky:2013ar,Brodsky:2013npa}  dilaton in AdS in order to simulate confinement in physical spacetime.  In that case, the parameter $\kappa=m_{\rho}/\sqrt{2}$ where $m_\rho$ is the mass of the $\rho$ meson. As discussed in reference \cite{Forshaw:2012im}, the normalization $\mathcal{N}_{\lambda}$ of the AdS/QCD wavefunction is fixed according to the polarization of the meson. 
The single free parameter in the AdS/QCD DAs is therefore the light quark mass $m_f$. In Ref. \cite{Forshaw:2012im}, $m_f$ was chosen as $0.14$ GeV because the AdS/QCD wavefunction 
was used in conjunction with a dipole model \cite{Soyez:2007kg} whose parameters were fitted to the HERA data on $F_2$ with $m_f=0.14$ GeV. This value for $m_f$ was also used in previous dipole model computations \cite{Forshaw:2011yj,Forshaw:2010py,Forshaw:2006np,Forshaw:2004vv} and also recently in Ref. \cite{Ahmady:2012dy}. 
In this paper, we shall also explore the possibility of using a constituent quark mass $m_f=0.35$ GeV and a current quark mass $m_f=0.05$ GeV. 
Finally, note that both DAs are normalized, i.e.
\begin{equation}
\int_0^1 \d z \; \phi_{\rho}^{\perp,\parallel} (z, \mu)=1 
\end{equation}
and that the decay constants are given by \cite{Forshaw:2003ki,Ahmady:2012dy}
\begin{equation}
f_\rho = \frac{N_c}{m_\rho \pi}  \int_0^1 \d z \;
\left.[z(1-z)m^{2}_{\rho} + m_{f}^2 -\nabla_{r}^{2}]
\frac{\phi_L(r,z)}{z(1-z)}
\right|_{r=0}
\label{vector-decay}
\end{equation}
 and
\begin{equation}
f_{\rho}^{\perp}(\mu) =\frac{m_f N_c}{\pi} \int_0^1 \d z \; \int \d r \; \mu J_1(\mu r)  \frac{\phi_T(r,z)}{z(1-z)} \;.
\label{tensor-decay-mu}
\end{equation}

In Table \ref{tab:couplings}, we compare the AdS/QCD predictions for the decay constants with those obtained using QCD sum rules and lattice QCD.  Note that we achieve better agreement with sum rules and the lattice when we use the constituent quark mass. However, a more favorable agreement with the datum on the experimentally measured $f_{\rho}$ is achieved with the lower quark masses. We should also recall that our AdS/QCD predictions for the scale-dependent decay constant $f_{\rho}^{\perp}$ hardly evolve with $\mu$ for $\mu \ge 1$ GeV, i.e. our predictions hold at a low scale $\mu$  of order $1$ GeV \cite{Ahmady:2012dy}. This is also the case for our AdS/QCD DAs \cite{Ahmady:2012dy,Forshaw:2012im}. 
\begin{table}[h]
\begin{center}
\[
\begin{array}
[c]{|c|c|c|c|c|c|}
\hline
\mbox{Reference}&\mbox{Approach} & \mbox{Scale}~\mu& f_{\rho} ~[\mbox{MeV}]&f_ {\rho}^{\perp} (\mu)~ [\mbox{MeV}] &f_\rho^{\perp}(\mu)/f_\rho \\ \hline
\mbox{\cite{Beringer:1900zz}}&\mbox{Experiment}& & 220 \pm 2 & & \\ \hline
\mbox{This paper}&\mbox{AdS/QCD} &\sim 1 ~ \mbox{GeV}&214,214,202&36, 95,152 &0.17, 0.45,0.75 \\ \hline
\mbox{\cite{Ball:2006eu}}&\mbox{Sum Rules} &2~\mbox{GeV} &206 \pm 7 &145 \pm 8 & 0.70 \pm 0.04\\ \hline
\mbox{\cite{Becirevic:2003pn}}&\mbox{Lattice} &2 ~ \mbox{GeV} & &&0.72 \pm 0.02 \\ \hline
\mbox{\cite{Braun:2003jg}}&\mbox{Lattice} &2 ~ \mbox{GeV} & &&0.742 \pm 0.014 \\ \hline
\end{array}
\]
\end{center}
\caption {AdS/QCD predictions for the decay constants of the $\rho$ meson compared to sum rules, lattice predictions and experiment. The three AdS/QCD predictions are given for $m_f=0.05,0.14$ and $0.35$ GeV respectively. }
\label{tab:couplings}
\end{table}

\begin{figure}
\centering
\subfigure[~The longitudinal twist-$2$ DA]{\includegraphics[width=.50\textwidth]{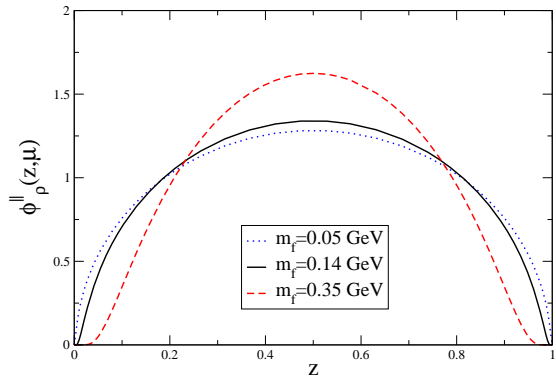} }
\subfigure[~The transverse twist-$2$ DA]{\includegraphics[width=.50\textwidth]{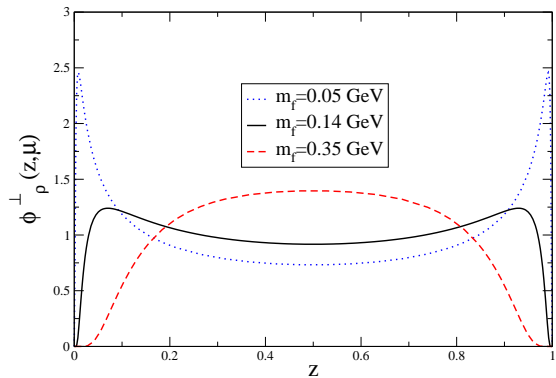} }
\caption{The twist-$2$ AdS/QCD DAs for three different quark masses: $m_f=0.05$ GeV (dotted blue), $m_f=0.14$ GeV (solid black) and $m_f=0.35$ GeV (dashed red).} \label{fig:DAs}
\end{figure}

The twist-$2$ AdS/QCD DAs are shown in Fig. \ref{fig:DAs}. As can be seen, both DAs widen as the quark masses decreases, with a remarkable end-point enhancement
in the transverse case.

\section{Form factors}
The matrix element for the radiative $B \to \rho \gamma$ decay is parametrized in terms of a single transition form factor, $F_1(q^2)$: 
\begin{equation}
\langle \rho, \lambda | \bar{d} \sigma_{\mu \nu} q^{\nu} b| B \rangle = i \epsilon_{\mu\nu\rho\sigma} e^{\nu *}_{\lambda} p_B^{\sigma} p_{\rho}^{\sigma} 2 F_1 (q^2) 
\label{F1}
\end{equation}
where $e_{\lambda}$ and $p_{\rho}$ are the polarization vector and $4$-momentum of the $\rho$ meson respectively. The $4$-momentum transfer to the photon is $q=(p_B-p_{\rho})$ where $p_B$ is the $4$-momentum of the $B$ meson. 

On the other hand, the matrix element for the semileptonic decay is parameterized in terms of 
 four form factors,  namely $A_{1,2,3} (q^2)$ and $V(q^2)$, i.e.
\begin{eqnarray}
\langle \rho,\lambda | \bar{u} \gamma^{\mu} (1-\gamma_5) b | B \rangle & = &
-i (m_B + m_\rho) A_1(q^2) \epsilon^{\mu *}_{\lambda} +
\frac{iA_2(q^2)}{m_B
+ m_\rho} (\epsilon^*_{\lambda} \cdot p_B) (p_B+p_\rho)^\mu\nonumber\\
& & {} + \frac{iA_3(q^2)}{m_B + m_\rho} (\epsilon^*_{\lambda} \cdot p_B)
q^\mu + \frac{2V(q^2)}{m_B + m_\rho}
\epsilon^\mu_{\phantom{\mu}\alpha\beta\gamma}\epsilon^{\alpha*}_{\lambda}
p_{B}^{\beta} p_{\rho}^{\gamma},\makebox[0.8cm]{}\label{eq:ME}
\end{eqnarray}
where $m_B$ is the mass of the $B$ meson and $q$ is now the $4$-momentum transfer to the lepton pair. The form factor $A_3$ does not contribute to the decay rate in the limit of vanishing lepton masses and we shall neglect it here. 

Note that for the radiative decay $B \to \rho \gamma$, $q^2=0$ in Eq. \eqref{F1}, but we shall be more general here and consider also non-zero $q^2$ which is relevant, for example, to the decay
$B \to \rho l^+ l^-$.  At the same time, this will allow us to test the Isgur-Wise relation \cite{Isgur:1990kf} as applied to the $\rho$ meson \cite{Ali:1993vd}: 
\begin{equation}
F_1^{\mbox{\tiny{IW}}}(q^2) = \left(\frac{q^2 + m_B^2 -m_{\rho}^2}{2 m_B} \right) \left(\frac{V(q^2)}{m_B + m_{\rho}} \right)+ \left(\frac{m_B + m_{\rho}}{2 m_B} \right) A_1(q^2) \;.
\label{IW}
\end{equation}
This relation, inspired by heavy quark symmetry and originally derived for $B \to K^*$ form factors,  is expected to become more accurate as $q^2$ increases to its maximum value $(m_B-m_\rho)^2 = 20.3~\mbox{GeV}^2$\cite{Ali:1993vd}. 

The above form factors can be computed using LCSR \cite{Ali:1993vd,Ball:1997rj,Ball:1998kk,Ball:2004rg}, lattice QCD \cite{Burford:1995fc,Flynn:1996rc,Lellouch:1996id,DelDebbio:1997kr} or quark models \cite{Scora:1995ty,Isgur:1988gb,Wirbel:1985ji,Faustov:1995bf,Jaus:1989au,Melikhov:1996pr}. In this paper, we shall use the LCSR in which the form factors are expressed in terms of DAs of the $\rho$ meson. The radiative form factor is given by \cite{Ali:1993vd}
\begin{eqnarray}
\lefteqn{F_1 (q^2)\ =\ \frac{m_b + m_f}{2 f_B m_B^2}\,
\exp\left\{
\frac{m_B^2-m_b^2}{M_B^2}\right\} \int_0^1
\frac{\d u}{u}\,\exp\left\{\frac{\bar
u}{uM_B^2}\,(q^2-m_b^2-um_\rho^2)\right\}}&&\nonumber\\
\displaystyle
&  \Theta[c(u,s_0^B)]\left\{
m_b f_{\rho}^{\perp} \phi_\perp(u,\mu) + u m_\rho
f_{\rho} g_\perp^{(v)}(u,\mu) + \left(\frac{m_b^2 + q^2 -u^2m_{\rho}^2 + uM_B}{4uM_B} \right) m_{\rho}f_{\rho} g_{\perp}^{(a)} (u,\mu) \right\} &\label{eq:LCF1}
\end{eqnarray}
while the semileptonic form factors are given by \cite{Ali:1993vd,Ball:1997rj}
\begin{eqnarray}
\lefteqn{A_1(q^2)\ =\ \frac{m_b + m_f}{f_B(m_B+m_\rho)m_B^2}\,
\exp\left\{
\frac{m_B^2-m_b^2}{M_B^2}\right\} \int_0^1
\frac{\d u}{u}\,\exp\left\{\frac{\bar
u}{uM_B^2}\,(q^2-m_b^2-um_\rho^2)\right\}}&&\nonumber\\
& & \Theta[c(u,s_0^B)]\left\{
 f_{\rho}^{\perp}\phi_\perp(u,\mu)\,
\frac{1}{2u}\,(m_b^2-q^2+u^2m_\rho^2) + m_b m_\rho
 f_{\rho} g_\perp^{(v)}(u,\mu)\right\}\!,\label{eq:LCA1}\\
\lefteqn{A_2(q^2)\ =\ \frac{(m_b + m_f)(m_B+m_\rho)}{f_Bm_B^2}\,\exp
\left\{
\frac{m_B^2-m_b^2}{M_B^2}\right\} \int_0^1
\frac{\d u}{u}\,\exp\left\{\frac{\bar
u}{uM_B^2}\,(q^2-m_b^2-um_\rho^2)\right\}}&&\nonumber\\
& & \left\{ \frac{1}{2}\, f_{\rho}^{\perp} \phi_\perp(u,\mu)
\Theta[c(u,s_0^B)]
+ m_b m_\rho f_{\rho} \Phi_\parallel(u,\mu) \left[ \frac{1}{uM_B^2}\,
\Theta[c(u,s_0^B)] + \delta[c(u,s_0^B)]\right]\right\}\!,\label{eq:LCA2}\\
\lefteqn{V(q^2)\ =\ \frac{(m_b+m_f)(m_B+m_\rho)}{2f_Bm_B^2}\,\exp\left\{
\frac{m_B^2-m_b^2}{M_B^2}\right\} \int_0^1
\frac{\d u}{u}\,\exp\left\{\frac{\bar
u}{uM_B^2}\,(q^2-m_b^2-um_\rho^2)\right\}}&&\nonumber\\
& & \left\{ f_{\rho}^{\perp} \phi_\perp(u,\mu) \Theta[c(u,s_0^B)]
+ \frac{1}{2}\,m_b m_\rho f_{\rho} g_\perp^{(a)}(u,\mu)\left[ 
\frac{1}{uM_B^2}\,
\Theta[c(u,s_0^B)] + \delta[c(u,s_0^B)]\right]\right\} \makebox[1cm]{\ }
\label{eq:LCV}\end{eqnarray}
with $c(u,s_0^B)=us_0^B-m_b^2 + q^2 \bar{u} - u \bar{u} m_{\rho}^2$ and where $\Phi_{\parallel}$ in Eq. \eqref{eq:LCA2} is given by
\begin{equation}
\Phi_\parallel(u,\mu) = \frac{1}{2}\,\left[ \bar u \int_0^u\!\!
\d v\,\frac{\phi_\parallel(v,\mu)}{\bar v} - u \int_u^1\!\!
\d v\,\frac{\phi_\parallel(v,\mu)}{v}\right] \;. 
\label{eq:bigphi}
\end{equation}
To leading twist-$2$ accuracy, the DAs $g_\perp^{v}$ and $g_\perp^{a}$ are given in terms of the twist-$2$ DA $\phi_{\parallel}$ \cite{Ball:1997rj}: 
\begin{equation}
g_{\perp}^{v}(u,\mu) = \frac{1}{2}\,\left[  \int_0^u\!\!
\d v\,\frac{\phi_\parallel(v,\mu)}{\bar v} + \int_u^1\!\!
\d v\,\frac{\phi_\parallel(v,\mu)}{v}\right] 
\label{eq:gvtw2}
\end{equation}
and
\begin{equation}
g_{\perp}^{a}(u,\mu) = 2 \left[  \bar{u}\int_0^u\!\!
\d v\,\frac{\phi_\parallel(v,\mu)}{\bar v} + u \int_u^1\!\!
\d v\,\frac{\phi_\parallel(v,\mu)}{v}\right] \;. 
\label{eq:gatw2}
\end{equation}

Therefore  the form factors depend on the twist-$2$ DAs of the $\rho$ meson as well as its decay constants $f_{\rho}^{\perp}$ and $f_{\rho}$ and also on parameters which characterize the $B$ meson channel, namely the Borel parameter $M_B$, the continuum threshold $s_0^B$, the quark mass $m_b $ and the $B$ meson decay constant $f_B$. Here we use the following set of parameter values : $M_B^2=6~\mbox{GeV}^2$, $s_0^B=34~\mbox{GeV}$ and $m_b=4.8~\mbox{GeV}$.   
We compute $f_B$ using the sum rule given in Ref. \cite{Ali:1993vd} in order to reduce the sensitivity of the form factors to the $b$-quark mass\cite{Ball:1997rj}. Note that the LCSR are more reliable at low and intermediate values of the momentum transfer $q^2$. On the other hand, lattice predictions are available at high $q^2$ and are thus complementary to LCSR predictions. 
Here, we shall extrapolate our predictions to the maximum value of $q^2$, i.e. $q^2 = 20.3~\mbox{GeV}^2$, in order to be able to compare to lattice predictions. 
 
\begin{figure}
\centering
\subfigure[~The semileptonic form factor $A_1$]{\includegraphics[width=.50\textwidth]{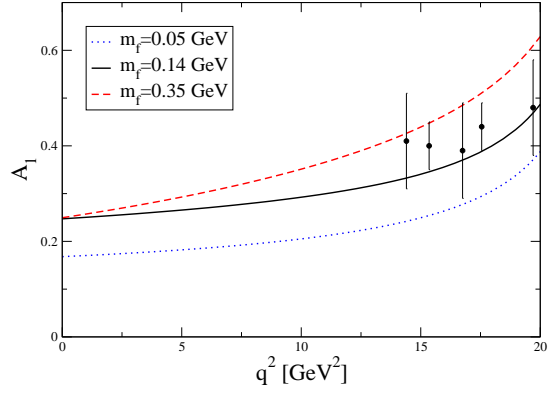} }
\subfigure[~The semileptonic form factor $A_2$]{\includegraphics[width=.50\textwidth]{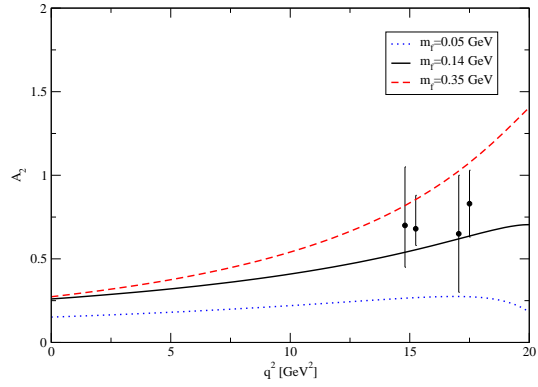} }
\subfigure[~The semileptonic form factor $V$]{\includegraphics[width=.50\textwidth]{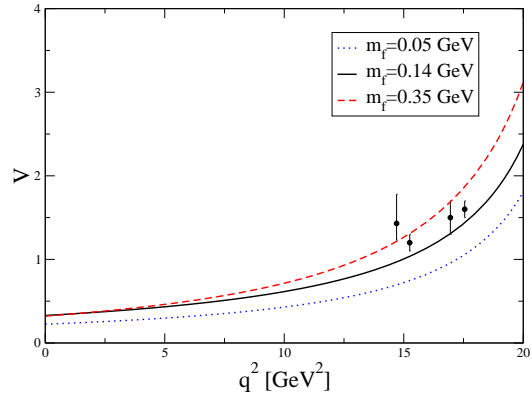} }
\caption{The semileptonic form factors $A_1$, $A_2$ and $V$ as functions of  $q^2$ for three different quark masses, $m_f=0.05,0.14$ and $0.35$ GeV. We extrapolate our predictions to high $q^2$ in order to compare to the lattice data from the UKQCD collaboration\cite{Flynn:1996rc,DelDebbio:1997kr,Burford:1995fc}.} \label{fig:semileptonicFF}
\end{figure}

\begin{figure}
\centering
{\includegraphics[width=.80\textwidth]{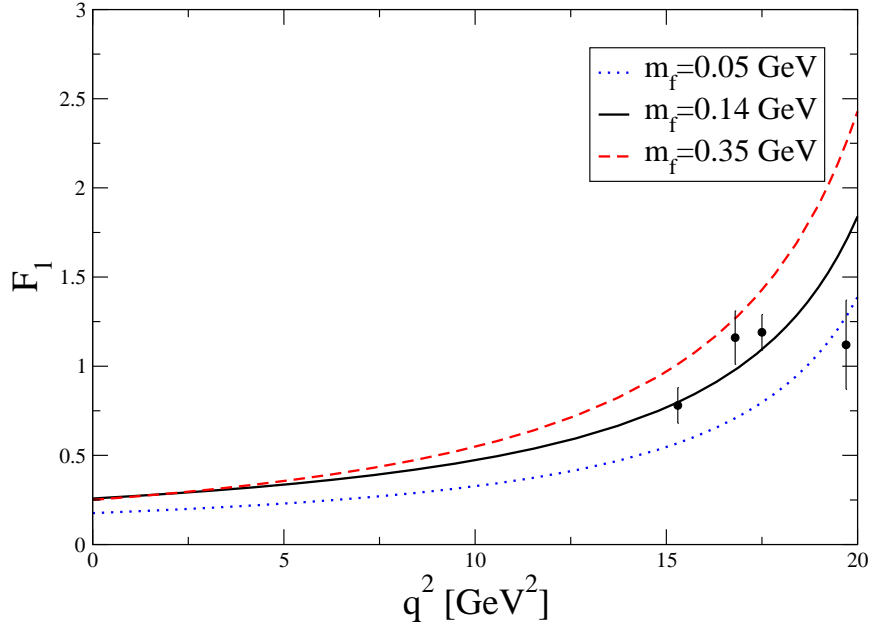} }
\caption{The radiative form factor $F_1$ as a function of $q^2$ for three different quark masses, $m_f=0.05,0.14$ and $0.35$ GeV. We extrapolate our predictions to high $q^2$ in order to compare to the lattice data from the UKQCD Collaboration \cite{DelDebbio:1997kr,Burford:1995fc}.}\label{fig:radiativeFF}
\end{figure}

Our predictions for the semileptonic and radiative form factors at $q^2=0$ are shown in Tables \ref{tab:semiFF} and \ref{tab:radFF} respectively.  As can be seen, a larger quark mass ($m_f=0.14$ GeV or $m_f=0.35$ GeV) yields better agreement with the predictions of  LCSR with sum rules DAs \cite{Ball:1997rj} and those of most quark models \cite{Faustov:1995bf,Wirbel:1985ji,Jaus:1989au}. In Fig. \ref{fig:semileptonicFF} and \ref{fig:radiativeFF}, we show the semileptonic and radiative form factors as a function of $q^2$ for our three different quark masses. We extrapolate our predictions beyond the region of reliability of the LCSR in order to be able to compare to lattice predictions. We can see that the heavier quark masses are preferred in order for our predictions to match the  lattice data. 

\begin{table}[h]
\begin{center}
\[
\begin{array}
[c]{|c|c|c|c|c|c|c|}\hline
\mbox{Form factor}  & \mbox{AdS/QCD}&\mbox{BB} &\mbox{FGM} & \mbox{WSB} &\mbox{Jaus}  &\mbox{Melikhov} \\ \hline
A_1(0) &0.17, 0.25,0.25 & 0.27 \pm 0.05&0.26 \pm 0.03 & 0.28 &0.26&0.17-0.18 \\ \hline
A_2(0) & 0.15,0.26,0.27 &0.28 \pm 0.05 &0.31 \pm 0.03 & 0.28 &0.24&0.155 \\ \hline
V(0) & 0.23,0.33,0.32 & 0.35 \pm 0.07 &0.29 \pm 0.03  & 0.33 &0.35& 0.215 \\ \hline
\end{array}
\]
\end{center}
\caption {Our predictions, corresponding to $m_f=0.05,0.14,0.35$ GeV, for the semileptonic form factors compared to the sum rules predictions of Ref. \cite{Ball:1997rj} and the quark model predictions of Ref. \cite{Faustov:1995bf,Wirbel:1985ji,Jaus:1989au,Melikhov:1996pr}.}
\label{tab:semiFF}
\end{table}

\begin{table}[h]
\begin{center}
\[
\begin{array}
[c]{|c|c|c|c|c|}\hline
\mbox{Form factor}  & \mbox{AdS/QCD} &\mbox{ABS}  & \mbox{BB} & \mbox{AOS} \\ \hline
F_1(0) & 0.18, 0.26,0.25 & 0.24 \pm 0.04 & 0.29 \pm 0.04 & 0.30 \pm 0.10\\ \hline
\end{array}
\]
\end{center}
\caption {Our predictions, corresponding to $m_f=0.05,0.14,0.35$ GeV, for the radiative form factor compared to the sum rules predictions of Refs. \cite{Ali:1993vd,Ball:1998kk,Aliev:1996hb}.}
\label{tab:radFF}
\end{table}

\begin{figure}
\centering
{\includegraphics[width=.80\textwidth]{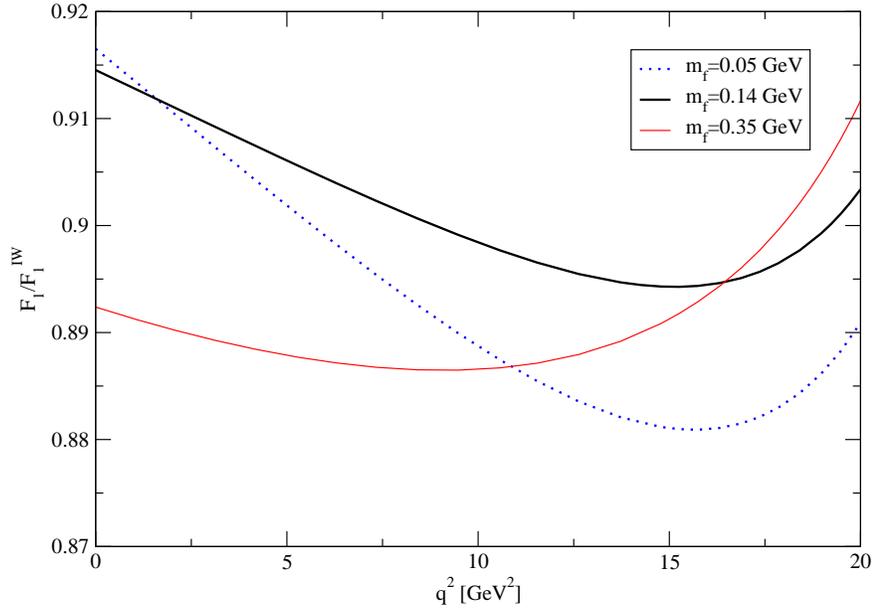} }
\caption{The ratio of the radiative form factor computed using LCSR to that computed using the Isgur-Wise relation.} \label{fig:F1}
\end{figure}

In Fig. \ref{fig:F1}, we show the ratio of the radiative form factor computed using Eq. \eqref{eq:LCF1} to that computed using the Isgur-Wise relation, Eq. \eqref{IW}, as a function of $q^2$. Again, we show predictions for our three different quark masses.  We observe that the Isgur-Wise relation is best satisfied for the 
larger quark mass, $m_f=0.35$ GeV, i.e. the ratio computed using this quark mass reaches values closer to unity at large $q^2$.  


\section{Semileptonic decay rates}
\begin{figure}
\centering
{\includegraphics[width=.80\textwidth]{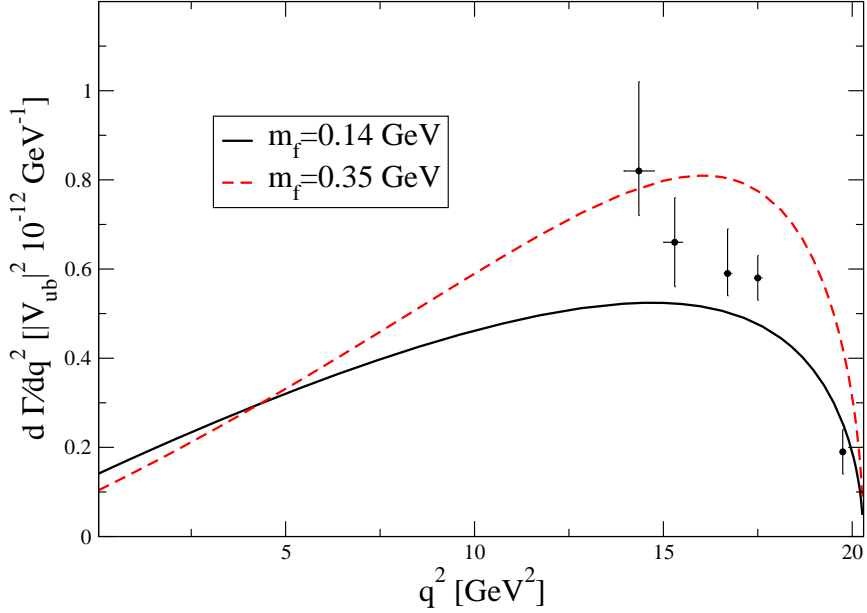} }
\caption{The $B\to \rho l \nu$ decay spectrum in $q^2$ computed using two different quark masses $m_f=0.14$ and $0.35$ GeV. Lattice data from the UKQCD collaboration \cite{Lellouch:1996id,DelDebbio:1997kr,Burford:1995fc}.} \label{fig:dGammadt}
\end{figure}

In this section, we compute the semileptonic decay rates.  Based on our observations in the previous sections, we shall exclude those predictions corresponding to the current quark mass, $m_f=0.05$ GeV. The transverse and longitudinal helicity amplitudes for the decay $B \to \rho l \nu$ are given by \cite{Ball:1997rj}
\begin{equation}
H_{\pm}(q^2)=(m_B + m_{\rho}) A_1(q^2) \mp \frac{\sqrt{\lambda(q^2)}}{m_B + m_\rho} V(q^2)
\end{equation}
and 
\begin{equation}
H_0 (q^2) = \frac{1}{2m_\rho\sqrt{q^2}}\left\{ (m_B^2 - m_\rho^2 -
t) (m_B + m_\rho) A_1(q^2) - \frac{\lambda (q^2)}{m_B +
m_\rho}\,A_2(q^2)\right \}
\end{equation}
respectively, where 
\begin{equation}
\lambda(q^2)=(m_B^2 + m_{\rho}^2 -q^2)^2 -4 m_B^2 m_\rho^2 \;.
\end{equation} 
The total differential decay width is then given by 
\begin{equation}
\frac{\d \Gamma}{\d q^2} =\frac{G_F^2|V_{ub}|^2}{192\pi^3m_B^3}\,\sqrt{\lambda(q^2)} \,q^2\left(H_0^2(q^2) +
H_+^2(q^2)+ H_-^2(q^2)\right).
\end{equation}
In Fig. \ref{fig:dGammadt}, we show the decay spectrum in $q^2$ for the two quark masses $m_f=0.14$ and $0.35$ GeV. As expected, our predictions are consistent with the lattice data at large $q^2$. 

By integrating over $q^2$, we obtain the total decay width which can be written as
\begin{equation}
\Gamma = \Gamma^L+ \Gamma^T
\end{equation}
where
\begin{equation}
\Gamma^L= \frac{G_F^2|V_{ub}|^2}{192\pi^3m_B^3}\, \int \d q^2 \sqrt{\lambda(q^2)} \,q^2\left(H_0^2(q^2) 
\right).
\end{equation}
is the decay width for longitudinally polarized $\rho$ mesons  and
\begin{equation}
\Gamma^T= \frac{G_F^2|V_{ub}|^2}{192\pi^3m_B^3}\, \int \d q^2 \sqrt{\lambda(q^2)} \,q^2\left(H_+^2(q^2) + H_-^2(q^2)
\right).
\end{equation}
is the decay width for transversely polarized $\rho$ mesons. Our predictions for the total decay width in units of $|V_{ub}|^2$ are compared to the LCSR calculation with sum rules DAs of Ref. \cite{Ball:1997rj} and with quark models \cite{Faustov:1995bf,Scora:1995ty,Jaus:1989au,Wirbel:1985ji,Melikhov:1996pr} predictions in Table \ref{tab:widths}. In Table \ref{tab:ratioLT}, we show predictions for the $|V_{ub}|$-independent ratio $\Gamma_L/\Gamma_T$.

\begin{table}[h]
\begin{center}
\[
\begin{array}
[c]{|c|c|c|c|c|c|c|c|}\hline
\mbox{Decay width}  & \mbox{AdS/QCD}& \mbox{BB} & \mbox{FGM} & \mbox{ISGW2} & \mbox{Jaus} &\mbox{WSB} & \mbox{Melikhov} \\ \hline
\Gamma/|V_{ub}|^2 & 12.0,15.9 & 13.5 \pm 4.0 & 5.4 \pm 1.2 & 14.2 & 19.1 & 26 & 9.64 \\ \hline
\end{array}
\]
\end{center}
\caption {Our predictions for the total decay width in units of $\mbox{ps}^{-1}$ computed using quark masses $m_f=0.14, 0.35$ GeV as compared to sum rules\cite{Ball:1997rj} and quark models \cite{Faustov:1995bf,Scora:1995ty,Jaus:1989au,Wirbel:1985ji,Melikhov:1996pr}.}
\label{tab:widths}
\end{table}

\begin{table}[h]
\begin{center}
\[
\begin{array}
[c]{|c|c|c|c|c|c|c|c|}\hline
\mbox{Ratio}  & \mbox{AdS/QCD}& \mbox{BB} & \mbox{FGM} & \mbox{ISGW2} & \mbox{Jaus} & \mbox{WSB} & \mbox{Melikhov} \\ \hline
\Gamma_{L} /\Gamma_{T}&0.59,0.42 &0.52 &0.5 \pm 0.3 & 0.3 & 0.82 & 1.34 & 1.13 \\ \hline
\end{array}
\]
\end{center}
\caption {Our predictions for the ratio of longitudinal to transverse decay widths using quark masses $m_f=0.14,0.35$ GeV compared to sum rules \cite{Ball:1997rj}  and quark models predictions \cite{Faustov:1995bf,Scora:1995ty,Jaus:1989au,Wirbel:1985ji,Melikhov:1996pr}. }
\label{tab:ratioLT}
\end{table}

Recently, the BaBar collaboration has measured partial decay widths in three different $q^2$ bins:\cite{delAmoSanchez:2010af}
\begin{equation}
\Delta \Gamma_{\mbox{\tiny{low}}}= \int_0^8 \frac{\d \Gamma}{\d q^2} \d q^2 = (0.564 \pm 0.166) \times 10^{-4} 
\end{equation}
for the low $q^2$ bin,
\begin{equation}
\Delta \Gamma_{\mbox{\tiny{mid}}}= \int_8^{16} \frac{\d \Gamma}{\d q^2} \d q^2 = (0.912 \pm 0.147) \times 10^{-4} 
\end{equation}
for the intermediate $q^2$ bin and
\begin{equation}
\Delta \Gamma_{\mbox{\tiny{high}}}= \int_{16}^{20.3} \frac{\d \Gamma}{\d q^2} \d q^2 = (0.268 \pm 0.062) \times 10^{-4} 
\end{equation}
for the high $q^2$ bin. 
From these measurements, we can thus deduce the $|V_{ub}|$-independent ratios of partial decay widths
\begin{equation}
R_{\mbox{\tiny{low}}}=\frac{\Gamma_{\mbox{\tiny{low}}}}{\Gamma_{\mbox{\tiny{mid}}}}=0.618 \pm 0.207
\end{equation}
and 
\begin{equation}
R_{\mbox{\tiny{high}}}=\frac{\Gamma_{\mbox{\tiny{high}}}}{\Gamma_{\mbox{\tiny{mid}}}}=0.294 \pm 0.083
\end{equation}
which we compare to our predictions: $R_{\mbox{\tiny{low}}}=0.580, 0.424$, $R_{\mbox{\tiny{high}}}=0.427,0.503$ for $m_f=0.14, 0.35$ GeV respectively.  Our predictions for $R_{\mbox{\tiny{low}}}$ are therefore in agreement with the BaBar measurement. This is not the case for $R_{\mbox{\tiny{high}}}$ where our predictions are above the BaBar measurement. This is perphaps not unexpected given that the LCSR predictions are less reliable in the high $q^2$ bin. 

\section{Conclusions}
We have predicted the radiative and semileptonic $B \to \rho$ form factors using light-cone sum rules with holographic AdS/QCD DAs and we have tested the Isgur-Wise relation between the various form factors.  We treated the light quark mass in the AdS/QCD DAs as a free parameter and  found that a quark mass between $0.14$ GeV and $0.35$ GeV is preferred. We also computed $|V_{ub}|$-independent observables for the semileptonic decay $B \to \rho l \nu$. Our predictions for the ratio of the partial decay width in the low $q^2$ bin to that in the intermediate $q^2$ bin are in agreement with the BaBar data. 
Our future goal is to compute the $B \to K^*$ form factors using the holographic AdS/QCD DAs for $K^*$ recently derived in Ref. \cite{Ahmady:2013cva}.

\section{Acknowledgements}
This research is supported by the Natural Sciences and Engineering Research Council of Canada (NSERC) and by the Provost\rq{}s Proposal Development Fund (PPDF) of Mount Allison University. 

\bibliographystyle{apsrev}
\bibliography{AdSformfactorv2_revised}

\end{document}